\documentclass[runningheads]{llncs}
\usepackage[linktoc=none, hidelinks]{hyperref}
\usepackage{graphicx}
\usepackage{xcolor}
\usepackage{soul}
\usepackage{booktabs}
\usepackage{tabularx}
\usepackage{enumitem}
\usepackage{tcolorbox}

\usepackage[framemethod=TikZ]{mdframed}

\usepackage{tcolorbox}
\tcbuselibrary{theorems}
\usepackage{fontawesome}

\newtcbtheorem{prompt}{\faWechat{} Prompt}
{colback=blue!5,colframe=blue!35!black,fonttitle=\bfseries}{th}

\newtcbtheorem{llmoutput}{\faWechat{} LLM Output}
{colback=blue!20,colframe=blue!35!black,fonttitle=\bfseries}{th}

\begin{document}
\title{Natural Language Processing for Requirements Traceability}

\author{Jin L.C. Guo\inst{1}
\orcidID{0000-0003-1782-1545} \and
Jan-Philipp Steghöfer\inst{2}\orcidID{0000-0003-1694-0972} \and
Andreas Vogelsang\inst{3}
\orcidID{0000-0003-1041-0815} \and
Jane Cleland-Huang\inst{4}\orcidID{0000-0001-9436-5606}}

\institute{McGill University, Canada \and
XITASO GmbH IT \& Software Solutions, Germany \and
University of Cologne, Germany \and
University of Notre Dame, USA}
\authorrunning{Guo et al.}
\maketitle             
\begin{abstract}
Traceability, the ability to trace relevant software artifacts to support reasoning about the quality of the software and its development process, plays a crucial role in requirements and software engineering, particularly for safety-critical systems. In this chapter, we provide a comprehensive overview of the representative tasks in requirement traceability for which natural language processing (NLP) and related techniques have made considerable progress in the past decade. We first present the definition of traceability in the context of requirements and the overall engineering process, as well as other important concepts related to traceability tasks. Then, we discuss two tasks in detail, including trace link recovery and trace link maintenance. We also introduce two other related tasks concerning when trace links are used in practical contexts. For each task, we explain the characteristics of the task, how it can be approached through NLP techniques, and how to design and conduct the experiment to demonstrate the performance of the NLP techniques. We further discuss practical considerations on how to effectively apply NLP techniques and assess their effectiveness regarding the data set collection, the metrics selection, and the role of humans when evaluating the NLP approaches. Overall, this chapter prepares the readers with the fundamental knowledge of designing automated traceability solutions enabled by NLP in practice.

\end{abstract}

\section{Fundamentals of Requirements Traceability}
\begin{figure}[t]
    \centering
    \includegraphics[width=0.7\textwidth]{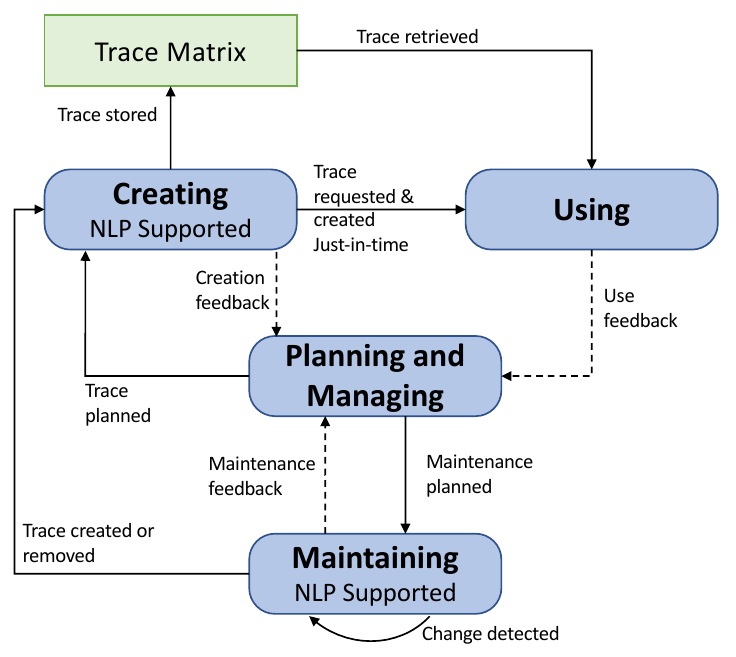}
    \caption{Requirements Traceability includes four fundamental activities of planning and managing the traceability strategy, creating links, maintaining links, and ultimately using links. NLP is particularly useful for supporting trace link creation and evolution.}
    \label{fig:trace-process}
\end{figure}

Requirements Traceability is defined by the Center of Excellence for Software and Systems Traceability (CoEST)\footnote{http://sarec.nd.edu/coest/} as the ability to link a requirement back to contributing sources such as stakeholders' rationales, hazards to be mitigated, and regulatory codes, and forward to corresponding design artifacts, code, and test cases.  A well-traced project serves as an indicator that a rigorous development process has been followed, and supports critical engineering activities such as requirements verification and validation, change impact analysis, safety analysis, software and systems compliance, and downstream activities such as regression testing.  Given these benefits, traceability is required in many regulated domains which often prescribe the types of software artifacts to be created and the traceability paths to be established. The traditional tracing process involves planning and managing a traceability strategy, creating and maintaining links, and finally using links as illustrated in Figure \ref{fig:trace-process}. 

In practice, the cost and effort of creating and maintaining trace links in a continually evolving software system is often perceived as excessively high, and therefore developers and other project stakeholders tend to avoid the overhead unless required by regulations. Several researchers have shown that even in regulated domains, trace links are often established in a relatively ad-hoc way, potentially as an afterthought for certification processes, leading to problems such as incomplete, incorrect, redundant, and even conflicting trace links.  Given all of these problems, traceability has traditionally been rarely used outside of regulated domains.

However, traceability can be so much more than this. A previously published Future of Software Engineering (FOSE) paper \cite{fose14} laid out the challenge of achieving \emph{ubiquitous traceability}. In this vision, traceability was fully integrated into the engineering process without any additional human effort needed to create or maintain trace links, and links were automatically generated and evolved through collecting, analyzing, and processing every piece of evidence from which trace data could be inferred and managed. Realizing this vision would mean that all projects across diverse domains, whether safety-critical or not, large or small, simple or complex, would benefit at no additional cost and effort from the presence of ubiquitous traceability.  This is particularly important given that the requirements landscape has changed dramatically over the past couple of decades. While some projects, especially those in the safety-critical domain still focus on traditional requirements processes, there are many other projects, in which traditional requirements are replaced by user stories or issues, and traceability, if it does exist, is, by necessity, more dynamic. 

This chapter lays out some of the cutting-edge work in the traceability domain that has sought to leverage Natural Language Processing (NLP) to see this vision become a reality. NLP-based tracing techniques can be used to augment a more traditional development approach by helping practitioners to construct and maintain trace links. They can also be used to dynamically generate just-in-time trace links as and when needed by practitioners.  As NLP techniques continue to evolve, supported by advanced Generative AI techniques such as Large Language Models (LLMs), the vision of ubiquitous traceability becomes increasingly plausible.

In the remainder of this chapter, we discuss a selection of important topics on applying NLP for traceability. We start with the most investigated task called trace link recovery (TLR), a task aiming to identify valid trace links from two sets of software engineering artifacts (Section~\ref{sec:link_recovery}). We then discuss how to maintain the validity of the trace links as the project evolves (Section~\ref{sec:link_maintenance}). Additionally, we direct the attention to two other often overlooked aspects of automated traceability solutions, including generating explanations on \textit{why} two artifacts are linked and classifying trace links to more fine-grained categories that better support project management (Section~\ref{sec:other_tasks}).

\section{Trace Link Recovery}
Over the past decades, numerous solutions have been proposed to meet the need for trace link recovery, ranging from popular information retrieval approaches to the latest Generative AI based approaches. This section discusses the representative methods from each line of work. 

\label{sec:link_recovery}
\subsection{Task Specification}
Trace link recovery (TLR) (or trace retrieval~\cite{fose14}) is the task of identifying traceability relations within an existing set of artifacts.  
Formally, the TLR task can be represented by two sets of artifacts: sources $S$ and targets $T$. TLR is the task to find a relation $R\subseteq S \times T$ such that $(s,t)\in R$ iff $s$ and $t$ are in a traceability relation. TLR can be conducted in two modes: (1) starting from \textit{one} source artifact, all related target artifacts shall be determined, or (2) a complete traceability model shall be recovered by recovering trace links for \textit{all} source and target artifacts.

TLR is particularly relevant in practice since trace links are often not created and maintained systematically from the beginning of a project or get lost over time. If traceability is needed later (e.g., for certification or impact analysis), TLR can help \textit{recover} the underlying traceability information.

TLR is probably the most studied task in the area of traceability. Zhao~et~al.\ conducted a systematic literature review (SLR) on using NLP for RE tasks and identified 53 publications focusing on TLR approaches~\cite{zhao2021natural}. TLR is adopted to relate artifacts from various activities in the development lifecycle. Pauzi~et~al.~\cite{pauzi2023applications} conducted an SLR specifically on NLP for software traceability and identified TLR approaches for a broad range of artifacts covering trace links between requirements and code, design artifacts, test cases, other requirements, and artifacts from operations. 

\subsection{Approaches}
\label{sec:link_recovery:approaches}
TLR approaches use techniques from \textit{information retrieval}, \textit{(shallow) machine learning}, or \textit{deep learning}. More recently, generative AI models have also been used to solve TLR by prompting LLMs. 

\subsubsection{Information Retrieval}

A good overview of TLR approaches that use methods from information retrieval (IR) is given in the systematic mapping study conducted by Borg~et~al.~\cite{borg2013recovering}.

The premise of IR-based approaches is that when two artifacts have a high degree of textual similarity, they should most likely be traced~\cite{oliveto2010equivalence}. Commonly used IR algorithms include Vector Space Models (VSM), Latent Semantic Indexing (LSI), and Latent Dirichlet Allocation (LDA).

\paragraph{Vector Space Models (VSM).}
Vector space models are algebraic models that represent documents and queries as vectors $d=(w_{1},\ldots,w_{N}), q=(q_1,\ldots,q_N)$. Each dimension $1,\ldots,N$ corresponds to a separate term of the entire corpus. If a term occurs in the document, its value in the vector is non-zero. 
The definition of \textit{term} depends on the application. Typically terms are single words, keywords, or longer phrases. If words are chosen to be the terms, the dimensionality of the vector is the number of words in the vocabulary (i.e., the number of distinct words occurring in the corpus).
Several different ways of computing these values, also known as weights or term weights, have been developed. Probably the simplest approach is \textit{bag-of-words}, where a vector dimension is set to 1 if and only if the corresponding term occurs in the current document. 

One of the best-known alternatives is \textit{term frequency-inverse document frequency (TF-IDF)} weighting, which has, for example, been used by Hayes~et~al.\cite{hayes2006advancing}. TF-IDF defines a product of local (specific for a document) and global (valid for the entire corpus) parameters. With TF-IDF, the value of dimension $i$ in the vector for document $d=(w_{1},\ldots,w_{N})$ is calculated by $w_i=\mathit{tf}_i(d)\cdot \mathit{idf}_i$, where $\mathit{tf}_i(d)$ is the (usually normalized) frequency of word $w_i$ in the document $d$ and $\mathit{idf}_i$, is computed as $\mathit{idf}_i=\log_2(\frac{n}{\mathit{df}_i})$ where $n$ is the number of documents in the corpus and $\mathit{df}_i$ is the number of documents in which word $w_i$ occurs. Intuitively, TF-IDF provides high values to terms that occur frequently in a specific document but rarely in other documents of the corpus. 

In vector space models, vector operations can be used to compare documents with queries (e.g., cosine similarity~\cite{hayes2006advancing} or Jaccard similarity~\cite{falessi2016estimating}). 

While TLR based on vector space models is relatively easy to implement, compute, and interpret, it has some severe drawbacks. First, vector space models are sensitive to exact word matching. Word substrings or synonyms used in queries will not be considered similar to words in the documents. This may be mitigated by preprocessing steps such as stemming, lemmatization, or stop word removal, or by including information from thesauri~\cite{hayes2006advancing}. Secondly, traditional VSMs neglect contextual information associated with a word such as the order of words or the specific surroundings of words. 

\paragraph{Latent Semantic Indexing (LSI).}
To overcome the challenge of dealing with synonyms and including contextual information, some approaches utilize Latent Semantic Indexing (LSI)~\cite{marcus2003recovering,hayes2006advancing}. LSI reduces the dimensions of the vector space, finding semi-dimensions using singular value decomposition. The new dimensions are no longer individual terms, but concepts represented as combinations of terms. The number of dimensions to which LSI reduces the vector space is a parameter, which can also be optimized based on existing data. In the overview by Borg~et~al.~\cite{borg2013recovering}, most LSI-based approaches reduced the number of dimensions to an absolute value between 10--20 or to a relative value between 20\% and 50\%.

\paragraph{Latent Dirichlet Allocation (LDA).}
As an alternative to algebraic models such as VSM and LSI, researchers have also tested statistical models from the field of IR~\cite{oliveto2010equivalence,asuncion2010software}. Latent Dirichlet Allocation (LDA) is a statistical model that clusters documents based on hidden latent variables (called \textit{topics}) in the underlying data. 
The input to LDA is a term-by-document matrix, which is derived from any vector space model. This $m\times n$ matrix assigns a value to each term and each document based, for example, on its TF-IDF value, where $m$ is the number of terms and $n$ is the number of documents. LDA takes this term-by-document matrix as input and generates as output a $k \times n$ matrix $\Theta$, called topic-by-document matrix, where $k$ is the number of topics. An entry $\Theta_{ij}$ of the topic-by-document matrix denotes the probability of the $j$th document
belonging to the $i$th topic. Since typically $k << m$, LDA is mapping the documents from the space of terms ($m$) into a smaller space of topics ($k$). The latent topics allow the clustering of documents based on their shared topics. More specifically, documents having the same relevant topics are grouped in the same cluster, and documents having different topics belong to different clusters.   

For example, in a requirements specification for autonomous driving, the terms pedestrian, rain, road, and construction site would suggest a \textit{driving context} theme, while the terms distracted, eyes, navigation, steering, and hands-off would suggest a \textit{driver-related} theme. There may be many more topics in the collection, e.g., topics related to car state, energy control, sensor availability, behavior, etc.\ that we do not discuss for simplicity's sake. Stop words in a language are usually filtered out by pre-processing before LDA is performed. Additionally, stemming or lemmatization might help improve the results of the LDA.
If the document collection is sufficiently large, LDA will discover such sets of terms (i.e., topics) based upon the co-occurrence of individual terms, though the task of assigning a meaningful label to an individual topic (i.e., that all the terms are \textit{driver-related}) is up to the user, and often requires specialized knowledge (e.g., for collection of technical documents). For TLR, however, it is not necessary to assign meaningful labels to the topics since the topics will only be used to recover trace links. This is done by comparing a document's distribution over topics $\Theta_d$ to determine the topical content of the document. For instance, if there are 4 topics and if $\Theta_d = [0.5, 0.1, 0.05, 0.35]$, then one can infer that document $d$ is mainly comprised of a combination of topics 1 and 4. Note that $\Theta_d$ can also be
compared against other $\Theta'_d$ (using a similarity measure such as Kullback-Leibler divergence or cosine distance) to obtain a ranked list of topically similar documents. 

Oliveto et al.~\cite{oliveto2010equivalence} have used LDA to recover trace links between use cases and classes in source code. Asuncion~et~al.~\cite{asuncion2010software} used LDA to develop a general search engine for all kinds of textual documents related to a project.

\subsubsection{(Shallow) Machine Learning}

With the breakthroughs that machine learning methods have achieved since the early 2010s, methods from machine learning (ML) have also been used to solve TLR and many other requirements engineering (RE) tasks. TLR is often framed as a classification task and solved by supervised learning methods~\cite{falessi2016estimating,mills2018automatic,rath2018traceability}. In contrast to the IR approaches mentioned above, where ground truth data is only needed to evaluate the performance of the approach, ML methods need a ground truth (both links and non-links) as an input to learn a classifier that is later used to predict trace links between unseen data.

\begin{figure}
    \centering
    \includegraphics[width=1\linewidth]{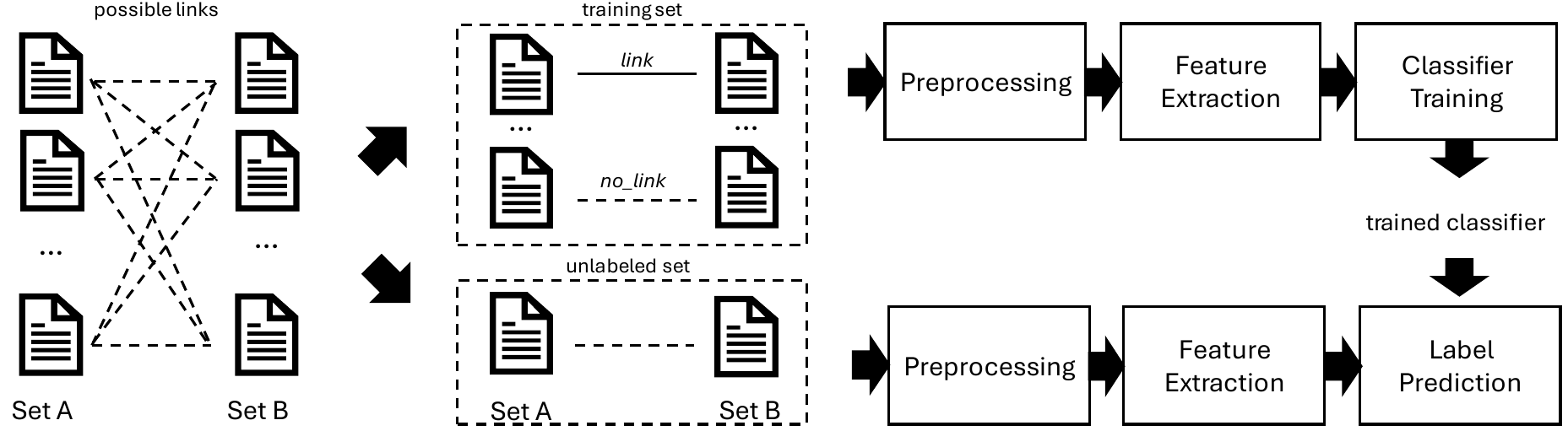}
    \caption{Schematic pipeline for an ML classifier for TLR.}
    \label{fig:classifier-pipeline}
\end{figure}

A typical pipeline for training and using an ML classifier is shown in Figure~\ref{fig:classifier-pipeline}.
The input for the pipeline is a dataset $T$ that consists of two sets $A$ and $B$ of documents with existing trace links between documents of the two sets. This dataset serves as a ground truth that is used to train and evaluate the classifier. As one of the first steps, the dataset needs to be split into disjoint sets that are used for training, testing, and validating the classifier. How the dataset is split influences implicit biases and information leakage, which may let the classifier look better than it is in a productive environment. A k-fold cross-validation strategy mitigates such a risk, where the dataset is randomly split into $k$ equally large folds of which $k-1$ folds are used to train the classifier and $1$ fold is used to test it. This procedure is run $k$ times such that each data point is tested exactly once. A more elaborate discussion about appropriate ways to split the data for training classifiers in SE has been published by Dell'Anna~et~al.~\cite{dellAnna2022evaluating}.   

The training data may then be preprocessed (e.g., stemming, lemmatization, stop word removal), and afterward transformed into a vector representation. This step is crucial since it defines the information that the learner can leverage. To calculate document vectors, researchers have used several techniques ranging from textual similarity~\cite{falessi2016estimating}, over more specialized features~\cite{falessi2016estimating}, or combinations of such~\cite{mills2018automatic,rath2018traceability}.

The classification task is then defined as: Given one artifact from the two sets in the training data, predict whether there exists a trace link between them or not. Thus, the task is a binary classification task with classes $\{\mathit{link}, \mathit{no\_link}\}$. Any pair of documents $d_1$ and $d_2$, such that $d_1\in A$ and $d_2\in B$, is assigned either $\mathit{link}$ or $\mathit{no\_link}$ as a label depending on whether there exists a trace link in the ground truth or not.

The resulting training set is usually very imbalanced because there are many more $\mathit{no\_link}$ than $\mathit{link}$ labels. Imbalanced data can bias the trained model to learn frequencies instead of input characteristics. Therefore, the training data set needs to be balanced to contain a similar number of document pairs with a link and without a link. This can be achieved by techniques such as undersampling or oversampling.  

To prepare the data for the training process, the vectors of the two input documents are concatenated and then fed into some ML algorithm suited for classification tasks (e.g., Naive Bayes~\cite{mills2018automatic,falessi2016estimating,rath2018traceability}, Logistical Regression~\cite{mills2018automatic}, Random Forest~\cite{mills2018automatic,rath2018traceability}, Support Vector Machines~\cite{mills2018automatic}).

To predict labels for unseen data, the learned model is fed with a pair of documents that have gone through the same preprocessing steps as the training data. The model then predicts a label $\mathit{link}$ or $\mathit{no\_link}$ for the document pair at hand.

\subsubsection{Deep Learning}
The performance of the aforementioned (shallow) machine learning approaches relies on how well the extracted features can capture the relevant semantic concepts and their relations. Deep learning offers better models to capture these semantics, especially by exploiting the context, in which a certain term is used. Therefore, the pipeline shown in Figure~\ref{fig:classifier-pipeline} is used similarly for deep learning approaches. Deep learning models can be used to transform a document into a richer and denser input vector or to serve as more advanced classification models.

Instead of using manually crafted features or simple text similarity measures, Guo~et~al.~\cite{guo2017semantically} used pretrained word embeddings as input representations for documents. Word embeddings are high-dimensional vector representations of words that have been learned by performing an unsupervised learning task on a corresponding text corpus. Popular word embedding models such as \textit{word2vec}, \textit{GloVe}, or \textit{fastText} have been trained on large sets of general-purpose texts and thus provide vector representations for words that reflect their meaning in documents quite well.  

Deep learning also offers new classification algorithms specifically suited to extract higher-level structures and patterns in textual data. Guo~et~al.~\cite{guo2017semantically} used Recurrent Neural Networks (RNN) as a classification model, which learns to predict the labels on the basis sequences of word embedding vectors. RNNs are specifically good at learning patterns in sequential data. Convolutional Neural Networks (CNN) is another deep learning classification model used for requirements classification~\cite{winkler2016automatic}, among the others. 

Pretrained word embeddings provided major advancements but had the problem that they could not reflect the specifics of certain technical domains and jargon used in requirements. Moreover, although word embeddings have been learned based on contextual use, they assign the same vector representation to a word independent of the context in which it is currently used. This limitation has led to the development of Large Language Models (LLMs) such as BERT~\cite{DBLP:conf/naacl/DevlinCLT19} that are trained on even larger amounts of data and can produce contextualized embeddings that are different depending on the context in which a word is used~\cite{LLMGuide}.

Lin~et~al.~\cite{lin2021traceability} were among the first to use the BERT model for TLR. BERT has a transformer architecture, which consists of two stages. The first is the creation of a foundation model, a language model for text in this case. This is done by unsupervised representation learning on a large set of documents related to the traceability task. Such foundation models can be reused for several tasks (not only TLR), and several pretrained models are already publicly available. In the second phase, the pretrained models are fine-tuned to the specific TLR task. This is done by continuing the training process with labeled data and adding a classification layer on top of the BERT architecture. 

\subsubsection{Generative AI}
Large language models have led to considerably better results in solving TLR as a classification task. In parallel, LLMs have been used generally for text generation and conversational systems. Systems like ChatGPT offer a linguistic interface to interact with and query LLMs through user-provided prompts. Rodriguez~et~al.~\cite{rodriguez2023prompts} tested the capability of Claude, a generative AI model, to solve TLR tasks directly by prompting it with instructions in natural language. 
They have shown that generative models can not only solve TLR tasks but can also be used to request more information about the reasons for the identified trace links. The authors further recognized the need for further research on how to craft effective prompts for TLR. The following is a concrete example provided by the authors for prompting the LLM to solve TLR:

\begin{prompt*}{}

Below are artifacts from the same software system. Is there a traceability link between (1) and (2)?

\vspace{0.2cm}
\noindent (1) The DPU-TMALI shall utilize SCM-DCI-SR, along with ERRNO provided by DPU-DCI to decode errors and place them on an error queue for DPU-CCM.

\vspace{0.2cm}
\noindent (2) Error Collection and Reporting: At boot time, no error queue exists because it has yet to be created.  Errors that occur in this early stage of error reporting are assigned directly to the global task variable errno.  If errno is set after the error queues are created, it is queued to the Error Queue by calling ccmErrEnq(). 

\end{prompt*}

\vspace{5pt}

\begin{llmoutput*}{}
Yes, there is a traceability link between (1) and (2). 
(1) mentions that the DPU-TMALI places errors on an error queue for the DPU-CCM.
\noindent (2) describes the error collection and reporting mechanism of the DPU-CCM, specifically referring to enqueuing errors.
\vspace{0.2cm}

\noindent So (2) provides an implementation detail for the error queuing functionality briefly mentioned in (1). Therefore, there is a traceability link between these two artifacts in the software system.
\end{llmoutput*}

\vspace{5pt}

One challenge of conversation-based LLMs is going to be their scalability. While having conversations about individual links, discussing hundreds of them with the generative AI is not time- or cost-efficient. In addition, the current approaches require the user to construct a prompt that includes the relevant artifacts. While this process can be automated, the sheer number of potential trace links in a reasonably sized trace data set would mean that there are thousands of prompts to the LLM, each incurring a small, but not negligible cost. Future work will have to address these scalability issues as well.

\begin{table}[t]
    \centering
    \setlength{\tabcolsep}{6pt}
    \setlength{\extrarowheight}{0.1cm}
    \caption{A context taxonomy of IR-based TLR evaluations. Level 1 is technology-oriented, and levels 3 and 4 are human-oriented. Level 2 typically has a mixed focus. Taken from \cite{borg2013recovering}.}
        \label{tbl:tlr-evaluation}
    \begin{tabularx}{\textwidth}{@{}p{3.9em}Xp{6.7em}p{7em}@{}}
        \toprule
        \textbf{Level 1:}\newline Retrieval context &  
        A strict retrieval context, performance is evaluated wrt.\ the accuracy of a set of search results. Quantitative studies dominate.& 
        Precision, \newline recall, \newline F-measure&
        Experiments on benchmarks, possibly with simulated feedback\\
        \textbf{Level 2:}\newline Seeking context &  
        A first step towards realistic applications of the tool. A seeking context with a focus on how the human finds relevant information in what was retrieved by the system. Quantitative studies dominate. &  
        MAP, DCG, Lag, DiffAR, DiffMR. & 
        Experiments on benchmarks, possibly with simulated feedback\\
        \textbf{Level 3:}\newline Work task context &  
        Humans complete real tasks but in an in-vitro setting. The goal of evaluation is to assess the causal effect of an IR tool when completing a task. A mix of quantitative and qualitative studies. &
        Time spent on task, \newline quality of work. & 
        Controlled experiments with human subjects.\\
        \textbf{Level 4:} \newline Project context & 
        Evaluations in a social-organizational context. The IR tool is studied when used by engineers within the full complexity of an in-vivo setting. Qualitative studies dominate. &
        User satisfaction, \newline tool usage &
        Case studies\\
        \bottomrule
    \end{tabularx}
\end{table}

\subsection{Evaluation}
\label{sec:link_recovery:evaluation}
Borg et al.~\cite{borg2013recovering} have identified four contexts in which TLR approaches can be evaluated (see Table~\ref{tbl:tlr-evaluation}). TLR is often a \textit{needle-in-the-haystack} problem, where a specific source artifact is compared to a large set of potential target artifacts of which only very few are related to the source artifact (i.e., traceability matrices are very sparse). For example, the NASA CM-1 dataset,\footnote{http://sarec.nd.edu/coest/datasets/CM1-NASA.zip} which is one of the most used public benchmarks for TLR, contains 235 high-level and 220 low-level requirements with only 361 true links. For this reason, several authors have stressed the importance of recall for TLR approaches. In general, the human-in-the-loop should always be considered when evaluating TLR approaches~\cite{maro2018vetting,cuddeback2010automated}.

Shin et al.~\cite{shin2015guidelines} performed a systematic literature review on evaluations for TLR approaches and defined three goals. Goal~1 is to find trace links with high accuracy, e.g., to support tasks like coverage analysis. Goal~2 is to find relevant documents excluding irrelevant documents to reduce the unnecessary effort for human analysts. Goal~3 is to rank all documents so that the relevant ones are near the top of the retrieved list, and also to reduce human effort.

\subsection{Limitations and Concluding Remarks}
\label{sec:link_recovery:limitations}
Although TLR is the traceability task that most work has been published on, existing approaches still have limitations.

\textbf{TLR approaches often do not give reasons or explanations for trace links.}
TLR approaches may have gone better in accurately predicting trace links over the last few years. Yet, effective use of these approaches relies on the capabilities of the human analyst to leverage the information recovered by the approach. A crucial point is whether the human analyst can understand \textit{why} two artifacts are traced to each other. Unfortunately, except for the Generative AI methods, other methods discussed above cannot provide explanations for the recovered links. So far, only a few approaches have addressed this important issue specifically (see Section~\ref{sec:link_explanation}).

\textbf{TLR approaches often do not distinguish different types of trace links.} Most TLR approaches recover trace links without any specific notion of traceability. However, traceability information may carry different semantics. For example, the relationship between two requirements may indicate that one requirement refines the other, conflicts with the other, is a duplicate of the other, or blocks the other. Detecting specific types of links is the first step toward understanding what information and why a TLR approach has recovered a link. However, only a few approaches currently address identifying different types of trace links. These will be further detailed in Section~\ref{sec:type_prediction}. 

\textbf{TLR approaches often lack large enough and credible datasets.} Laliberte~et~al.~\cite{laliberte2022evaluation} draw a dire conclusion: according to them, ``natural language processing is likely not a practical approach to requirements traceability.'' The state of the art does indeed show the limitations of NLP for trace link recovery and the lack of ``big data'' which could be used to train large models is an inherent property of the problem~\cite{guo2017semantically}. A lack of large, open datasets with an agreed-upon ground truth of links also poses a problem for researchers. Nevertheless, the research community has continuously chipped away at the issue and novel ideas such as using LLMs for TLR have the potential to reinvigorate the field.

\section{Trace Link Maintenance}
\label{sec:link_maintenance}

Once traceability links have been established, they need to be kept up-to-date and maintained throughout the development lifecycle. For this purpose, some approaches that use NLP are known and will be described in the following sections.

\subsection{Task Specification}
\label{sec:link_maintenance:specification}

Trace link maintenance (TLM) is defined as ``all activities associated with keeping traceability links up to date and consistent''~\cite{maro2016traceability}. It is necessary to devote effort to this task since traceability information is subject to decay as the information that is being connected via traceability links changes~\cite{mader2012towards}. Such changes happen in the normal course of the development of a system. In general, a traceability model needs to evolve along with the system under development. Whenever changes are made to the development artifacts, there is a chance that the traceability links need to be updated as well.

In this chapter, we will focus on automated ways of addressing trace link maintenance with NLP. Batot~et~al.\ categorize trace link maintenance as well as the related notion of trace link integrity as part of the larger category of trace management~\cite{batot2022survey}. In their taxonomy, we consider the approaches presented here as being automated link vetting approaches, even though the vetting process does not necessarily include the notion of automated fixes of the trace links being vetted.

Arguably, any practical trace link maintenance approach needs to be able to deal with a mixture of automatically generated and manually generated trace links and leave the manually created ones untouched. If this requirement is fulfilled, even a flawed trace link recovery mechanism can be used to generate an initial trace matrix which is then refined by manual work from experienced engineers. Once the original matrix has been created, trace link recovery can be used to feed a vetting process, in which a tool supports the engineers by making suggestions, which can be manually accepted or rejected (see, e.g., \cite{maro2018vetting}).

This kind of approach would at least address new links that need to be added during development.
In practice, there are a number of ways how existing traceability links need to be updated to ensure consistency. As a traceability link is a semantic relationship between at least two development artifacts, updates can either a) change which artifacts are related to each other, b) change the semantics of the relationship, or c) remove the traceability link completely.

\paragraph{Change which artifacts are related to each other.}
A typical example of a change that makes it necessary to change the artifacts that are related to each other is \emph{refactoring the system architecture}. Such a change might make it necessary to update the traceability links between components or classes in the architecture description and the respective source code files. This particular problem has been addressed in the past by using, e.g., a tool that identifies edit operations on a UML class diagram and makes the necessary changes in the trace matrix automatically~\cite{mader2009enabling}.

Another example of changes in the related artifacts is \emph{requirements refinement}. For instance, a larger requirement is split up and a part of it is moved into a new, separate requirement. Traceability links between the original requirement and the tests that were used to validate the externalized part of the requirement should then be updated.

\paragraph{Change the semantics of the relationship.}
In ``semantically rich traceability spaces''~\cite{maro2016traceability}, traceability links can carry a number of semantics that describe the relationships between the artifacts. Such information is often captured in \emph{traceability information models} whose design is difficult and can contain ambiguities~\cite{steghofer2021design}. In particular, they can contain semantic ambiguities. For instance, a requirement can \texttt{reference} another requirement or \texttt{refine} another requirement. It is either up to the engineer to decide which semantics to assign to a traceability link or the decision might have been made by an automated approach. In any case, the type of relationship between artifacts can change over time. Using the example of \emph{requirements refinement} again, a requirement that used to reference another one can be rewritten to become a refinement, thus necessitating updating the semantics.

If a traceability model carries little semantic information, e.g., if the traceability links are implicit and established via naming conventions or IDs, this scenario is less relevant -- since there is no distinction of the different types of relationships between artifacts, a link either exists or does not and updates are performed by removing traceability links.

\paragraph{Remove the traceability link.}
In other cases, existing traceability links lose their value and should be removed. One possibility is that a system refactoring has moved some functionality out of a component and into a different one so that the trace links connecting the requirements to the functionality in the original component can be removed. A scenario that also sometimes occurs in practice is that engineers identify that a link had been created erroneously and that the artifacts the link connects are not in a relationship.

\subsection{Approaches}
\label{sec:link_maintenance:approaches}

In general, trace link recovery approaches are far more common in the literature than maintenance approaches. However, the boundaries between these approaches are also not sharp: most of the approaches described in Section~\ref{sec:link_recovery:approaches} can also be used in trace link maintenance, at least to some degree. As Maro et al.~\cite{maro2016traceability} note, if trace recovery was perfect, the links could be recreated from scratch every time the artifacts have changed. The current trace matrix is then simply replaced by a newly generated one. This would make explicit trace maintenance unnecessary.
Of course, this way of doing things can incur significant processing times, makes it impossible to combine automated and manual tracing, and generally is only advisable if the performance of the approach is sufficient. In practice, none of the automated trace recovery approaches are sufficiently good to be useful when dealing with real-world, safety-critical systems~\cite{maro2018software}. 

Rahimi and Cleland-Huang propose an approach that combines trace link recovery and trace link maintenance~\cite{rahimi2018evolving}. Their \emph{Trace Link Evolver} detects changes in the linked artifacts and identifies a specific change scenario. Such a scenario includes typical refactorings, like merged classes, moving methods between classes, or making functionality obsolete. Depending on the scenario, links are then either newly created or removed. As changes are detected incrementally, this approach does not recover the entire trace matrix at once, but rather creates and removes individual traces.
NLP is used specifically to process requirements and source code to determine if new functionality has been added. Both kinds of artifacts are pre-processed to remove stop words and perform stemming. They are then fed into a Vector Space Model (VSM) which uses TF-IDF to assign an importance value to each word in the artifact (see also in Section~\ref{sec:link_recovery:approaches}). Cosine similarity is then used to compare requirements and source code. If the similarity is above a certain threshold and no trace link currently exists, one of the change scenarios applies and a new link between the requirement and the class or the method is created.

Mills et al.~\cite{mills2018automatic} motivate their \emph{TRAIL (TRAceability lInk cLassifier)} approach by explicitly referring to the challenge of keeping traceability links up-to-date and of discovering new trace links in the changing development artifacts. The principle of their approach is to train a machine learning classifier that distinguishes true links from false ones in the set of all possible links between the artifacts. The training of the classifier happens on the existing trace links -- the approach therefore learns from the trace links that have already been identified as correct. The classifier uses a number of features that are constructed using NLP approaches. For instance, one feature is constructed using a VSM in which the importance of the terms in an artifact is stored using TF-IDF and two artifacts can be compared using cosine similarity. Another example is LDA (see also in Section~\ref{sec:link_recovery:approaches}) in which an artifact is represented as a vector of probabilities indicating if a term is present in the artifact. The Hellinger distance can be used to compute similarities between artifacts based on this model. In all cases, common preprocessing steps were applied to the texts, such as removing underscores or Java keywords and stemming.

Arunthavanathan et al.~\cite{arunthavanathan2016support} use a number of NLP tools to translate requirements into the structured format used by Stanford CoreNLP\footnote{\url{https://stanfordnlp.github.io/CoreNLP/}} which differentiates, e.g., noun phrases and verb phrases. This helps to identify design concepts in the requirements that can then be found in UML class diagrams. The \emph{Software Artefacts Traceability Analyzer (SAT-Analyzer)} normalizes pronouns used in the requirements text (e.g., replacing all instances of ``they'' with ``the client''), employs stemming and morphological analysis to find the root forms and finally also employs WordNet\footnote{\url{https://wordnet.princeton.edu/}} to identify synonyms, hyponyms, and hypernyms. Based on the standardized representation of the requirement, the tool then uses a rule-based approach to identify classes, methods, and attributes in the requirements. These elements are added to an internal dictionary. Source code in Java and class diagrams in XMI format are also parsed in a similar way. The constructs found therein are matched to the internal dictionary. If a match is found, a new trace link is created.

Updates to any of the artifacts are propagated to the internal dictionary and the identified trace links. This keeps the trace links up-to-date and also means that the tool creates new trace links if, e.g., a requirement has been updated and new keywords were introduced that are also found in the source code or the class diagrams. If the tool detects that an artifact has been deleted, it also deletes the corresponding trace links.

\subsection{Evaluation}
\label{sec:link_maintenance:evaluation}

Maro et al.~\cite{maro2016traceability} describe the notion of a consistency function that maps all development artifacts and the trace links in the trace matrix to a value between 0 and 1. Such a function can be used to check if the overall consistency of the traceability links improves or declines during development. A traceability maintenance approach should thus be ``consistency improving'' to ensure that the quality of the trace links does not deteriorate over time~\cite{maro2016traceability}.

The problem is, of course, to define a suitable consistency function. Maro et al.~provide an example of a consistency function that combines validity (i.e., the trace links conform to the defined structure of the traceability information model), completeness (percentage of artifacts connected with at least one traceability link), and correctness (the traceability links that do exist are actually correct). While the first two aspects can be determined automatically, the second one is more difficult. In the case of Maro et al.~\cite{maro2016traceability}, the authors argue that this step needs to be performed manually.

In practice, the approaches reported in the literature simplify this significantly by using the same metrics that are also used to evaluate trace recovery approaches (cf.~Section~\ref{sec:link_recovery:evaluation}). Arunthavanathan et al.~\cite{arunthavanathan2016support} as well as Mills et al.~\cite{mills2018automatic} and Rahimi and Cleland-Huang~\cite{rahimi2018evolving} use precision, recall, and F$_1$ measure. Only the latter justifies why more complex metrics like Lag~\cite{DBLP:journals/tse/HayesDS06} or MAP~\cite{DBLP:conf/sigsoft/LoharAZC13} were not chosen: the approach is a binary classifier (should there be a trace link or not?) and does not create rankings. This is a limitation of all three approaches described here and shows that the limitations described in Section~\ref{sec:link_recovery:limitations} also apply to trace link maintenance.

Interestingly, Arunthavanathan et al.~\cite{arunthavanathan2016support} as well as Mills et al.~\cite{mills2018automatic} evaluate their approaches exactly as trace link recovery approaches are evaluated. They evaluate exactly one version of a software system and use typical metrics from TLR. While Rahimi and Cleland-Huang~\cite{rahimi2018evolving} evaluate different versions of the system, they also use precision and recall. While this gives an overall picture of the quality of the approach, it does not consider the changes between versions. A consistency-improving TLM approach as defined by Maro et al.~\cite{maro2016traceability} would leave correct links in place, add new links that are necessary, and update or remove existing links as needed. Precision and recall do not measure this as they do not compare two solutions with each other but only operate on distinct versions of the artifacts and trace links.

As a further note, not all papers compare to a baseline. Only Rahimi and Cleland-Huang use VSM and LSI as baselines~\cite{rahimi2018evolving} to show that their approach is superior to established ones.

\subsection{Datasets}
\label{sec:link_maintenance:datasets}

Arunthavanathan et al.~\cite{arunthavanathan2016support} mention four different datasets, namely ``bank requirement'', ``order requirement'', ``hostel management system'', and ``coach tour management system'' but do not share any information about them. In particular, they do not describe the number of requirements, classes, etc., and do not provide any details about how the ground truth has been created. It is not clear either if any artifact updates were involved in the evaluation. The datasets are not available to the public either.

Mills et al.~\cite{mills2018automatic} use datasets from COEST\footnote{\url{http://sarec.nd.edu/coest/datasets.html}}, in particular eAnci, EasyClinic, eTour, iTrust, MODIS, and SMOS. These datasets contain a number of different artifact types and also provide a ground truth for the trace links. However, even though Mills et al.\ emphasize the importance of maintenance in their paper repeatedly, they also do not evaluate the performance of their approach in maintenance tasks, but rather only report on the results for trace link recovery.

Rahimi and Cleland-Huang~\cite{rahimi2018evolving} use a custom dataset created by having developers evolve two different applications. They thus create several evolved versions that contain a number of refactoring for each app which allows them to effectively evaluate the \emph{Trace Link Evolver}, a tool they proposed. The ground truth (i.e., traceability links existed before and after the changes) was created by the researchers. TLE was then used to detect the changes and to update the traceability links. TLE's results were then compared to the ground truth to calculate precision, recall, and F$_{2}$-measure.

In general, it is even more difficult to construct a ground truth for trace link maintenance than it is for trace link recovery. The reason is that each version of an app requires its own ground truth. Rahimi and Cleland-Huang manually constructed this ground truth for the rather large changes performed by the developers based on an analysis of the source code and a description of the changes that were also provided by the developers.

However, ideally, each change that a TLM approach would look at is relatively small and localized. Guidelines for developers usually indicate that they should commit small, individual changes~\cite{purushothaman2005towards}. Analyzing such small changes for maintenance actions that modify the trace links is a simpler approach. Indeed, Rath et al.~\cite{rath2018traceability} exploit this for their approach to augment incomplete trace links. Likewise, Mukelabai et al.~\cite{mukelabai2023featracer} use commits to automate their feature location approach. A traceability maintenance mechanism could do the same thing -- the construction of the ground truth, namely which trace links should change, would then be more localized and easier to manage for each commit.

\subsection{Limitations and Concluding Remarks}
\label{sec:link-maintenance:limitations}

Many of the limitations of TLR mentioned in Section~\ref{sec:link_recovery:limitations} also hold for trace link maintenance. In terms of explainability, only the approach by Rahimi and Cleland-Huang provides justifications for changes to the trace matrix -- the change scenarios provide a clear indication of why a trace link was established or removed. However, all approaches discussed here are unable to differentiate between different semantic relationships -- as discussed in Section~\ref{sec:link_maintenance:evaluation}, they all act as binary classifiers (either there should be a link or there should not) and are not able to assign more meaningful semantics. In principle, it is possible to predict the link type by using multi-class classification (cf.~Section~\ref{sec:type_prediction}), but this has not been applied to trace link maintenance yet. Finally, as discussed in Section~\ref{sec:link_maintenance:datasets}, constructing meaningful datasets for trace link maintenance that cover several changes of a system and ideally even the entire development lifecycle while providing a ground truth for each step is a formidable challenge.

In addition to this, trace link maintenance brings its own, unique challenges:

\textbf{TLM approaches often do not consider manual changes.} The few trace link recovery approaches described in the literature do not make explicit accommodations for manual changes to the trace matrix, but instead operate under the assumption, that engineers will fully rely on the automated approach. In particular, it is not clear if these approaches allow trace links to be protected from removal or modification or if they allow for the use of the information gathered in a vetting process. In practice, however, engineers should be able to manipulate the trace matrix alongside an automated approach without their changes being overridden.

\textbf{TLM approaches often do not change a link, but remove it and create new links.} One of the maintenance operations identified above is to change a link in terms of which artifacts are connected. The TLM approaches described here all ``update'' a link by removing the old link and creating a new one. This is difficult since trace links can carry additional information\,--\, apart from semantic information, they can carry comments about why they were created, who created them, information about their history, and other things. In particular in domains in which information needs to be audited and accountability is important, such information cannot be lost (see, e.g., \cite{steghofer2021design}). Updating an existing link also improves the traceability of the trace matrix itself, especially if it is versioned appropriately.

\section{Other Tasks}
\label{sec:other_tasks}

The implication of natural language processing techniques on traceability goes beyond trace link construction and maintenance. Considering concrete contexts where trace links are used, many other tasks related to traceability can directly benefit from the advantages of adopting NLP methods. In this section, we introduce two such tasks: trace link explanation and trace link type prediction.

\subsection{Trace Link Explanation}
\label{sec:link_explanation}

\subsubsection{Task Specification}
Trace link explanation is the task of facilitating the understanding of trace links by explaining why two artifacts are linked. Trace links are often cross two different types of artifacts (e.g., requirements, design document, source code, etc.). Depending on the nature of the artifacts, the vocabulary used, and the domain knowledge embedded, the connection of the two artifacts tracing to each other can be obscure. While automated methods described in Section~\ref{sec:link_recovery} can handle such a gap~\cite{guo2017tackling} to a large extent, they are far from perfect. Requirement analysts or experts still have to go through the vetting process to confirm the existence of the links returned by those methods.   In such a case, explicitly explaining the rationales of the trace link is beneficial for them to effectively evaluate the automatically retrieved links. For stakeholders with limited domain knowledge, such explanations can help them properly interpret and understand the trace links. Moreover, for safety and security-critical systems where traceability is required, the explanation of links can also support the auditing process. 

The barriers to understanding a trace link can appear at different levels. On the artifact level, the terminology and acronyms appearing in individual artifacts can be unfamiliar to readers, interfering with their understanding of the semantics of the artifact. For example, to understand the following regulatory requirement specified by the USA Certification Commission for Health Information Technology (CCHIT): ``\textit{The system shall provide the ability to retrieve, display, store, and export a HITSP C/48 document}'', analysts have to properly interpret, among others, what ``HITSP'' stands for (i.e., Healthcare Information Technology Standards Panel). 

Furthermore, on the trace link level, connecting two artifacts often requires contextualizing the domain concepts that appeared in each artifact and reasoning their relationship. Properly generating the explanation at each level requires not only processing the artifacts through language technology but also representing and locating relevant domain knowledge. The previous requirement is linked to the following requirement from the US Veteran’s WorldVista healthcare system:
``\textit{The system shall have the capability to capture and store risk, social, and medical factors for each new patient.}''. The link can only be understood when the analyst grasps how the ability to operate on \textit{HITSP C/48 document} is related to storing the medical record of the patients. 

In sum, two related tasks fall under the scope of trace link explanation. 
\begin{enumerate} [topsep=0pt, partopsep=0pt, itemsep=0pt, parsep=0pt]
    \item Given an artifact, identify and explain the domain-specific terms appearing in the artifact.
    \item Given a trace link between two artifacts, explain relationship between concepts appearing in both artifacts.
\end{enumerate}

\subsubsection{Approaches}
Given that trace link explanation can be divided into two related but distinct tasks, approaches for addressing trace link explanation also fall into two categories, i.e. explaining terms and explaining relations. 

\vspace{3mm}
\noindent \textbf{Explaining terms}
The first step for explaining terms is to identify them from the artifacts. For artifacts written in natural language, syntactic analysis methods are often used, such as Part-of-Speech (POS) tagging and dependency parsing, to identify the noun phrases that are likely to be domain-specific concepts and need explanation. The CoreNLP tool~\footnote{https://stanfordnlp.github.io/CoreNLP/} by StanfordNLP group is widely adopted for those tasks. 

Many previous works aim at explaining domain-specific concepts in the artifacts for the purpose of supporting comprehension, mostly relying on external resources such as Wikipedia~\cite{8306825,10098113}, project glossaries~\cite{10.1145/3510003.3510129}, and domain corpus~\cite{10.1145/3510003.3510129}. 

Structured resources such as Wikipedia, in particular, the knowledge base built from Wikipedia entries and their connections, can effectively supplement the information of the artifacts if the knowledge base contains entries of the relevant concepts. Locating the entry in Wikipedia or similar knowledge bases that correspond to the terms mentioned in a natural language artifact is often called \textit{wikification}. Nassif and Robbilard have performed an in-depth analysis of the performance of existing wikification tools, also known as \textit{wikifiers}, on software artifacts~\cite{nassif2021wikifying}. For example, using JSI Wikifier,\footnote{https://wikifier.org/} one of the wikifiers examined in this work, the following concept would be identified using the default setting:

\vspace{1mm}
The \underline{system} shall provide the ability to retrieve, \underline{display}, \underline{store}, and export a \underline{HITSP} C/48 document.
\vspace{1mm}

Each underlined text is annotated by the wikifier with the Wikipedia entry in which the term appears. In this case, the page of ``American National Standards Institute'' is linked to the term ``HITSP''. When the term appears in multiple possible entries, a ranked list of entries is returned based on the calculation of confidence (JSI Wikifier uses the pagerank value). Note that the wikification process itself involves identifying noun phrases in the artifacts and evaluating the relationship between them and the entries in the knowledge base. They are well encapsulated within the wikifiers so that the users can use them as a black box to perform both term identification and explanation.

The relevance of the annotation provided by a wikifier can vary greatly. Some of the terms are too general, making the explanation redundant, e.g., ``system'' in the previous example. A predefined list can be effective in such cases to filter frequent but general terms. Liu et al.\ created such a blacklist by comparing the general corpus with the domain-specific ones and identifying the overlapped concepts~\cite{10.1145/3510003.3510129}. Moreover, some project-specific terms might not be mentioned in the knowledge base at all. In this case, a project glossary can provide more precise explanations for the target terms. Liu et al., for example, leveraged a glossary from the WordVista project~\footnote{https://tinyurl.com/WorldVistaGlossary} in which key acronyms and concepts for this domain are explained.

In cases where the term cannot be found in either the general knowledge base or the project glossary, online research offers a viable alternative. For example, Liu et al.\ suggested that a query following the template  ``what is inbody:$<$concept$>$ in $<$domain name$>$'' returns consistently relevant results to build the domain-specific corpus using the Bing search engine~\cite{10.1145/3510003.3510129}. They then used the syntactic analysis methods mentioned above to find expressions in the corpus that are likely to be explanations, e.g., in the form of ``$<$concept$>$ is/are/do something''. To filter out expressions outside the scope of the target project domains, they trained a binary classifier to evaluate if the concept itself, the definition, or the context sentence belongs to the target project domain.

\vspace{3mm}
\noindent \textbf{Explaining relations}
Various relationships might appear in a trace link that can help explain why the link should be established. Decisions have to be made on what kind of relationship to provide and for what purpose. Liu et al.\ focused on the hierarchical or equivalent relations between domain concepts appearing in two artifacts. This is achieved by first finding the $\langle Subject, Verb, Object \rangle$ triplet from the domain corpus while confining verbs representing a hierarchical (e.g., includes) or equivalent (e.g., equals) relationship. A knowledge graph is then formed by all the identified triplets, with the nodes representing the domain concepts and edges indicating their relationship. To explain the links, the shortest path in the knowledge graph between two concepts from the artifacts is returned, if it exists. For the example above, the shortest path would include: \newline $\langle HITSP C\/48, contains, Medical Summary Document\rangle$, \newline $\langle Medical Summary Document, contains, Allergy Concern Entry\rangle$, \newline $\langle Allergy, is \: a, Risk\rangle$

Guo et al. approached link explanation through more extensive syntactic analysis and a predefined set of heuristics~\cite{guo2015trace}. They proposed a syntactic template called \textit{Action Frame} that is centered around a verb in the artifacts (therefore action). Other than the verb, each action frame is characterized by roles such as agent (i.e.,  subject of the verb), theme (i.e., object of the verb), etc.  One or more action frames can be extracted from each artifact to represent the transition of the systems' states. For example, one action frame can be populated by analyzing this requirement for the patient-controlled analgesia (PCA) pump software: ``The \underline{upstream monitor} [Agent] measures drug flow into the pump and \underline{detect}s [Action] \underline{upstream occlusion} [Theme].''  Then, by applying the heuristics to the action frames from two artifacts, rationales of the trace link can be generated. For example, the rationale for the basic rule is formulated as 
\texttt{`Both artifacts involve' + G(Agent1, Agent2) + Action1 + `ing' + G(Theme1, Theme2)}, in which \texttt{RoleName1} indicates the role extracted from the source artifact, \texttt{RoleName2} is the role extracted from the target artifact, and    \texttt{G(RoleName1, RoleName2)} in the pattern represents the more general concept between the contents in \texttt{RoleName1} and \texttt{RoleName2} as defined in the domain knowledge graph. Clearly, generating such an explanation needs not only a knowledge graph for domain concepts but also a set of heuristics that defines the action frame mapping rules and rationale generation template.

\subsubsection{Evaluation}
When evaluating the task of trace link explanation, both aspects of verification and validation should be considered. First, we care about the coverage and accuracy of the methods at each step. For example, research questions can be asked concerning the domain concept identification step, such as how many concepts are identified from the artifacts, what percentage of the identified concepts are domain-specific, and how many domain concepts in the artifacts are missing. Similar questions can be asked about the concept explanation and relationship explanation. Given that we seldom have ready-to-use ground truth data for such information, manual annotation is unavoidable, which might pose concerns about the quality and size of the evaluation dataset. 

For validation, researchers often ask questions such as whether the generated explanations are useful for analysts for certain tasks. These types of questions can be answered by designing and conducting user studies with analysts or a similar population, during which the explanations are presented to them for a particular context. While the study normally requires more effort to carry out, often further involving developing prototypes with UI components for the tool, they can help derive insightful outcomes about the analysts and how they might use the trace link explanation in practice~\cite{10.1145/3510003.3510129}.

\subsection{Link Type Prediction}
\label{sec:type_prediction}

\subsubsection{Task Specification}
Similar to trace link explanation, link type prediction is a task that goes one step further from predicting whether a link exists between two artifacts\,--\,it aims to provide richer information about the links, in particular, predicting if the reason for creating the trace links falls into a certain category. When the notion of software traceability was first proposed, it was often applied to specific artifacts for safety-critical software, such as faults identified from failure analysis, requirements specifications, test cases, etc. The types of trace links are normally easy to infer from the types of artifacts. For example, a test case \textit{tests} a requirement specification, or a requirement specification \textit{mitigates} an identified fault. The link type can be clearly specified through a Traceability Information Model (TIM)~\cite{mader2013strategic}. Given the rise of open source software development and their common adoption of modern issue tracking systems such as GitHub\footnote{https://github.com/} and Jira,\footnote{https://www.atlassian.com/software/jira} the types of artifacts under consideration are much more heterogeneous, and the trace links types are therefore much less well defined. For example, Nicholson and Guo~\cite{nicholson2021issue} revealed that 12 different labels were assigned to issues of the three open source projects in Jira, such as bug, tasks, improvement, sub-tasks, new features. The frequently appeared link types between those issues are \textit{related to}, \textit{duplicates}, \textit{blocks}, \textit{depends upon}, \textit{requires}, etc. Such practical usage of trace links between issues calls for a more precise prediction of the trace links. Consequently, the trace link type prediction task can be specified as a multi-class classification task -- predicting the label from a predefined set of labels indicating the link type of two artifacts under evaluation.

\subsubsection{Approaches}
Given the availability of large amounts of existing data from issue trackers such as Jira~\cite{montgomery2022alternative},  link type prediction is often approached through learning-based methods. The first step in those methods is to represent the two involved artifacts as a vector for training the classifier and later making inferences. Each issue in the issue tracker is characterized by its title, a more detailed description and other metadata, such as reporter, created time, and issue type. 
Researchers need to decide on what information to use to encode for individual issues and how. For example, Nicholson and Guo~\cite{nicholson2021issue} encoded both the textual data (title and description, using TF-IDF or fastText embeddings trained by Wikipedia, StackOverflow, or project-specific documents) and the metadata (issue type, reporter identifier and assignee identifier using one-hot encoding). 
L{\"u}ders et al.~\cite{9920034}, on the other hand, chose to encode issues only using their title and description (through general BERT model~\footnote{https://github.com/google-research/bert}) considering those are universal features across different issue trackers. Then, a pair of artifacts can be represented as a combination of the features of individual artifacts, with optionally other features on the relations of the two involved artifacts (such as differences in creation time between the two issues). When using BERT to encode the textual content from the pair of artifacts, either the [CLS] token~\footnote{[CLS] stands for classification and is a special token added at the beginning of the input sequence. The vector of [CLS] token is considered as approximating the sequence representation since [CLS] token is used when performing the next sentence prediction (NSP) objective (either the second sentence is the next one in sequence) during pre-training.} or averaging over the sequence can be used.
For predicting the type of the link, machine learning methods introduced in Section~\ref{sec:link_recovery:approaches} are generally applicable to the vector representation of the artifact pair, such as Logistic Regression, Random Forest, or Neural Network.

\subsubsection{Evaluation}
While the methods for link type prediction resemble the ones used for link recovery, additional care is needed when designing the experiment to evaluate the performance of the methods, considering the characteristics of the data. First of all, the quality of the dataset needs to be carefully scrutinized before being used for training or testing the machine learning models. Previous works have pointed out that the use of link labels in Jira is  inconsistent~\cite{nicholson2021issue,montgomery2022alternative,9920034}. For example, multiple terms with slight variations are used as link labels in Jira, such as \textit{Depend}, \textit{Dependency}, \textit{Dependent}, \textit{Depends}. In such a case, it might be advisable to unify those labels. On the other hand, some of the labels with similar meanings might be better left without merging. Considering the label \textit{Clone}, an indication of duplication, it might be better to be considered as different from the label \textit{Duplicate} since the two labels are often assigned in different contexts. While \textit{Duplicate} is assigned by users explicitly after both involved issues are created, the label \textit{clone} is automatically created by Jira when its users use the ``clone'' feature to create new issues from existing ones. Through a qualitative labelling process, Montgomery et al.~\cite{montgomery2022alternative} investigated initial data mined from Jira and created a cleaned dataset with around 1 million issue links. 

The second consideration is how to address the class imbalance issue. Naturally occurring labels are heavily skewed towards common types such as \textit{Related to},  \textit{Duplicate}, and \textit{Subtasks}, etc. The number of instances observed for the minority classes, such as \textit{Cause} and \textit{Requires}, might be insufficient to train the classifier, which might lead to inferior overall performance. Methods frequently used to handle class imbalance include class weights and SMOTE~\cite{chawla2002smote}. Unfortunately, no decisive improvement can be observed in previous studies when applying those methods~\cite{nicholson2021issue,9920034}. A closely related consideration is what the appropriate metrics are when comparing different methods or methods with different options to configure. Metrics such as macro, micro, and weighted F$_1$ are all related to aggregating the F$_1$ measure across all classes. The key difference is in how they consider the class imbalance issue. Macro F$_1$ considers each class equally -- once the F$_1$ measure is calculated for each class, their average is used to indicate the performance of the classifier. Micro F$_1$, on the other hand, considers each instance equally and therefore ignores the factors of imbalanced data -- the true positives, false negatives and false positives across all classes are first aggregated, using which F$_1$ is then calculated. Finally, weighted F$_1$ calculates the class level F$_1$ first; then, it is averaged while weighing the frequency of label occurrences. The selection of the metrics should be suitable for the assumptions of the importance of different classes and the intended use cases of the classifier. 

The last factor is time since it plays a critical role when interpreting how the methods might perform when deployed in real systems, during which only historical data is observable and can be used for training the machine learning models. The training data used in the experiment, therefore, should resemble the historical data when deployed. The models might be applied to make predictions on other historical data, e.g., the links overlooked by developers. In such a case, a random split of the experimental data into training and testing would be justifiable. When the dataset is not big enough, cross-validation is often used to overcome the impact of randomness. In the case when the models need to make predictions on future events, a timestamp-based data split is more appropriate. The older experimental data can be used for training, while the newer data can be used for testing.

\section{Summary and Conclusion}

As illustrated in Figure~\ref{fig:trace-process}, traceability covers fundamental activities concerning the planning and managing of traceability strategies, creating and maintaining links, and supporting the use of links in context. This chapter provides an overview of how the advances in NLP have helped with some of those activities. Over the past decades, most has been devoted to the trace link recovery task. Other aspects, such as trace link maintenance and link type prediction, have also attracted notable attention. More future progress calls for novel ways to collect or create high-quality trace datasets that contain information on fine-grained categories of link types and how they evolve along with the software project. As hinted by the latest work on Generative AI, the conversational interface backed by large language models serves as an intuitive way for trace link creation and explanation, but it is not yet clear how these approaches will scale. Their ability to integrate project and work context opens new opportunities for designing automated traceability solutions toward higher levels of relevance. At the same time, more work is needed on how to design and improve the reliability of the Generative AI-based methods and systematically evaluate their impact. This chapter also uncovers the limited progress in using NLP in traceability planning. Realizing the vision of ``ubiquitous traceability'' calls for more effort in supporting early stages of the engineering process and considering traceability as part of the project priority. NLP methods might help better allocate manual and automated resources towards more effective traceability solutions.

\bibliographystyle{splncs04}
\bibliography{chapter}
\end{document}